\documentclass[sigconf]{acmart}

\usepackage{multirow}
\usepackage{makecell}
\usepackage{esvect}
\usepackage{threeparttable}
\usepackage{enumitem}

\usepackage[font=small,skip=0.5pt]{caption}
\usepackage{amsmath,bm}
\usepackage{graphicx}
\usepackage{caption}
\usepackage{subfigure}

\usepackage{subfloat}
\usepackage{booktabs}
\usepackage{multirow}
\usepackage{color}
\usepackage{algorithm}
\usepackage{algorithmic}
\usepackage[switch]{{lineno}}
\usepackage{colortbl}
\definecolor{mygray}{gray}{.9}
\usepackage{dsfont}
\usepackage{bbding}
\usepackage{pifont}
\usepackage{wasysym}
\usepackage{footmisc} 

\AtBeginDocument{%
  \providecommand\BibTeX{{%
    \normalfont B\kern-0.5em{\scshape i\kern-0.25em b}\kern-0.8em\TeX}}}

\setcopyright{acmcopyright}
\copyrightyear{2024}
\acmYear{2024}
\acmDOI{XXXXXXX.XXXXXXX}

\acmConference[WWW'24]{the ACM Web Conference 2024}{May 13--17, 2024}{Singapore}
\acmPrice{15.00}
\acmISBN{978-1-4503-XXXX-X/18/06}

\begin{document}

\title{On Practical Diversified Recommendation with Controllable Category Diversity Framework}

\author{Tao Zhang}
\email{selous.zt@alibaba-inc.com}
\affiliation{%
  \institution{Alibaba Group}
  \city{Hangzhou}
  \state{Zhejiang}
  \country{China}
}
\author{Luwei Yang}
\authornote{Corresponding author.}
\email{luwei.ylw@alibaba-inc.com}
\affiliation{%
  \institution{Alibaba Group}
  \city{Hangzhou}
  \state{Zhejiang}
  \country{China}
}
\author{Zhibo Xiao}
\email{xiaozhibo.xzb@alibaba-inc.com}
\affiliation{%
  \institution{Alibaba Group}
  \city{Hangzhou}
  \state{Zhejiang}
  \country{China}
}
\author{Wen Jiang}
\email{wen.jiangw@alibaba-inc.com}
\affiliation{%
  \institution{Alibaba Group}
  \city{Hangzhou}
  \state{Zhejiang}
  \country{China}
}
\author{Wei Ning}
\email{wei.ningw@alibaba-inc.com}
\affiliation{%
  \institution{Alibaba Group}
  \city{Hangzhou}
  \state{Zhejiang}
  \country{China}
}
\renewcommand{\shortauthors}{Tao Zhang, et al.}

\begin{abstract}
 Recommender systems have made significant strides in various industries, primarily driven by extensive efforts to enhance recommendation accuracy. However, this pursuit of accuracy has inadvertently given rise to echo chamber/filter bubble effects. Especially in industry, it could impair user's experiences and prevent user from accessing a wider range of items. 
 One of the solutions is to take diversity into account. However, most of existing works focus on user's explicit preferences, while rarely exploring user's non-interaction preferences. These neglected non-interaction preferences are especially important for broadening user's interests in alleviating echo chamber/filter bubble effects.
 Therefore, in this paper, we first define diversity as two distinct definitions, i.e., user-explicit diversity (U-diversity) and user-item non-interaction diversity (N-diversity) based on user historical behaviors. Then, we propose a succinct and effective method, named as Controllable Category Diversity Framework (CCDF) to achieve both high U-diversity and N-diversity simultaneously.
 Specifically, CCDF consists of two stages, User-Category Matching and Constrained Item Matching. The User-Category Matching utilizes the DeepU2C model 
 and a combined loss to capture user's preferences in categories, and then selects the top-$K$ categories with a controllable parameter $K$.
 These top-$K$ categories will be used as trigger information in Constrained Item Matching. 
 Offline experimental results show that our proposed DeepU2C outperforms state-of-the-art diversity-oriented methods, especially on N-diversity task. The whole framework is validated in a real-world production environment by conducting online A/B testing. The improved conversion rate and diversity metrics demonstrate the superiority of our proposed framework in industrial applications.
Further analysis supports the complementary effects between recommendation and search that diversified recommendation is able to effectively help users to discover new needs, and then inspire them to refine their demands in search.
\end{abstract}





\maketitle

\begin{figure}[t]
	\centering
	\includegraphics[width=1.0\linewidth]{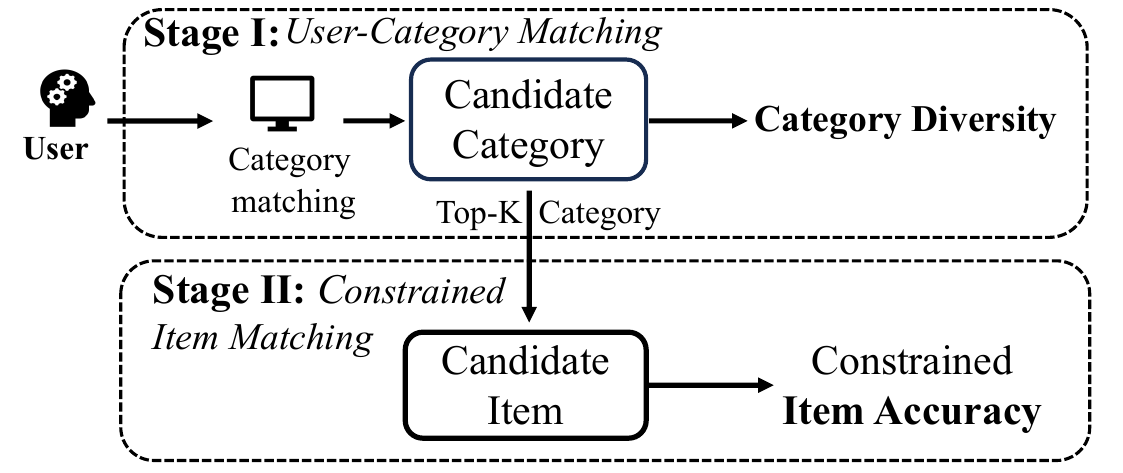}
	\caption{The proposed Controllable Category Diversity Framework (CCDF), which is divided into two stages. Top: User-Category Matching. Bottom: Constrained Item Matching}
	\label{fig:motivation}
\end{figure}
\section{Introduction}
Recommender systems (RS) have achieved great success in many industries, e.g., e-commerce~\cite{DBLP:journals/internet/LindenSY03}, online news~\cite{DBLP:conf/kdd/WuWAHHX19}, and multimedia content~\cite{DBLP:conf/sigir/ChenZ0NLC17}. Pursuing the accuracy of recommendation based on user attributes and historical preferences has attracted tremendous attention~\cite{Cheng2016WDL,Zhou2018DIN,Zhou2019DIEN,Xiao2020DMIN,tan2021sparse}. However, when accuracy is the only target, echo chamber/filter bubble effects will inevitably be introduced in recommendation~\cite{pariser2011filter, Ge2020EchoChamber, Yang2023DGRec}. It means user's interests will be reinforced by exposed items which are recommended by learning from user's historical preferences. As a result, the user will be isolated in this small set of items from his/her interests. This phenomenon will hurt user's experience and prevent user from accessing more diverse merchandise in e-commerce platforms or videos in video sharing websites.


To alleviate the echo chamber/filter bubble effects, diversity in recommender systems has attracted a growing research interest~\cite{wu2022survey}. The diversified recommendation is proven to help users to broaden their interests and improve their experiences~\cite{KUNAVER2017154}. Some post-processing techniques,  Maximal Marginal Relevance (MMR)~\cite{DBLP:conf/sigir/CarbonellG98} and Determinantal Point Process (DPP)~\cite{DBLP:conf/cikm/WilhelmRBJCG18}, focus on reshuffling and re-ranking the retrieved candidates to promote diversity. However, these methods are applied after candidate generation stage.
In order to consider diversity in candidate generation stage, some Graph Neural Network based methods are proposed~\cite{Zheng2021DGCN,Ye2021DDGraph,Yang2023DGRec,DBLP:conf/sigir/0001DWLZ020}.
They aim to expand the candidate pool by leveraging graph structures or content-based similarities, which expands the range of items available for recommendation. 
However, all these works evaluate diversity based on the number of categories, ignoring the relationship between users and categories. Even within diversified recommendations,
there exists the possibility that the presented categories are still within user's historical interactions (U-diversity).
The key to alleviating the echo chamber/filter bubble effects is to provide users with new choices (N-diversity), rather than just diversifying among user's existing interests.
The detailed definitions of U-diversity and N-diversity can be found in Section~\ref{sec:diversity}.

Therefore, in this paper, we propose a succinct and effective method, called Controllable Category Diversity Framework (CCDF)~\footnote{The code will be released at https://github.com/selous123/CCDF}, which is able to explicitly control category diversity and achieve both high U-diversity and N-diversity. The pipeline of the framework is shown in Figure~\ref{fig:motivation}. This framework is divided into two stages. The first stage is User-Category Matching, which is responsible for  selecting top-$K$ categories that the user may be interested in. 
The second stage is Constrained Item Matching, where the emphasis lies on accurately retrieving items within the constraints of the top-$K$ categories.
This process divides the responsibilities of category diversity and item accuracy into two different stages, User-Category Matching and Constrained Item Matching respectively, which equips CCDF with the controllability on category diversity by parameter $K$.
If $K$ is small, the retrieved items will be more related to user's historical preferences. If $K$ is large, the retrieved items will be more diversified. As a result, the user can access more diversified and novel items with larger $K$, thereby alleviating echo chamber/filter bubble effects.

Specifically, we novelly formalize User-Category Matching as a next category prediction task and construct the DeepU2C model, which employs User Net, Category Net and Wide Layer, to predict the next category label based on user historical behaviors. Moreover, we propose a combined loss function to further improve the performance.
We evaluate our DeepU2C model on U-diversity and N-diversity tasks to show its effectiveness in category diversity modeling.
To further validate the effectiveness of the CCDF, we conduct online experiments on DeepU2C model in conjunction with Constrained Item Matching which utilizes posterior weighted score. The online A/B testing results demonstrate significant improvements in both conversion rate and diversity metrics,
confirming the effectiveness of our proposed CCDF in industrial applications.
Further analysis supports the complementary effects~\cite{yuan2022recommendation} between recommendation and search that diversified recommendation is able to effectively help users to discover new needs, and then inspire them to refine their demands in search.

Our main contributions are summarized as follows:
\begin{itemize}
        \item We innovatively define the diversity as two distinct definitions, user-explicit diversity (U-diversity) and user-item non-interaction diversity (N-diversity), to evaluate the effectiveness of our method more comprehensively.
	\item We propose a succinct and effective two-stage Controllable Category Diversity Framework (CCDF), which is able to control category diversity explicitly and achieve both high U-diversity and N-diversity simultaneously.
	\item We formalize User-Category Matching as a next category prediction task, and then utilize the DeepU2C model and a combined loss to capture user's preferences in categories. These categories are used as trigger information for Constrained Item Matching.
	\item We compare our method on offline real-world industrial datasets with state-of-the-art methods. We also implement it in a real-world industrial production environment for A/B testing. The results validate the superiority of our method. The emerging complementary effects between recommendation and search are observed.
\end{itemize}

\section{Related Work}
\subsection{Accuracy in Recommendation} 
As one of the primary objectives of recommendation, accuracy has been studied extensively since it was proposed. The goal is to accurately recommend items based on user's preferences or interests. In industry, the recommendation system architecture is usually a funnel architecture. 
The first funnel is match module (aka candidate generation module) which is responsible for retrieving thousands of items from a large candidate pool. Collaborative Filtering~\cite{DBLP:journals/internet/LindenSY03} is one of the most representative match techniques which has been widely applied in industry. In recent years, deep learning techniques have been utilized to generating candidates. Deep candidate generation model~\cite{covington2016deep} applies deep learning techniques to effectively assimilate many signals and model their interactions. Some followed works, e.g., MIND~\cite{li2019multi} and ComiRec~\cite{cen2020controllable}, introduce effective modifications to the user and item deep learning networks and negative sampling techniques to improve the accuracy of user and item matching.
The second funnel is rank module which is used to rank the retrieved items elaborately. There are numerous research papers on rank module. Some of them are applied successfully in industry, such as WDL~\cite{Cheng2016WDL}, DeepFM~\cite{Guo2017DeepFM}, DIN~\cite{Zhou2018DIN}, DIEN~\cite{Zhou2019DIEN} and DMIN~\cite{Xiao2020DMIN}.
Almost all of the methods in match and rank modules emphasize on the accuracy of recommended items based on user's preferences or interests. 
Nevertheless, a sole emphasis on accuracy may inadvertently result in a dearth of diversity in recommendation which may cause users to fall into echo chamber/filter bubble effects~\cite{Ge2020EchoChamber, Yang2023DGRec}.

\subsection{Diversity in Recommendation} Diversity is introduced to solve the echo chamber/filter bubble effects in order to make the recommended results more plentiful.
Most earlier methods achieve diversity by re-ranking the recommendation list based on the user-item relevance scores and some diversity metrics. Maximum Marginal Relevance (MMR) ~\cite{DBLP:conf/sigir/CarbonellG98} is one of the most representative diversity methods in this field, which considers a linear combination of the relevance and diversity scores. It greedily selects the item if it has both high relevance score to the user and high diverse score to previously selected items.
Determinantal Point Process (DPP)~\cite{DBLP:conf/cikm/WilhelmRBJCG18} directly models the dissimilarity among items in a set-wise way instead of pair-wise way which has been used in MMR. A unified model is used to capture the global correlations among items.
In addition to the diversity approach on re-ranking stage, a number of graph-based methodologies~\cite{Zheng2021DGCN,Ye2021DDGraph,Yang2023DGRec} have been proposed in match module to directly generate diverse items. These techniques create a heterogeneous graph by conceptualizing users and items as nodes and abstracting behavioral interactions as edges. The intuitive sense that the graph-based methods are able to improve the diversity is that higher order neighbors of a user contain more diverse items.
DGCN~\cite{Zheng2021DGCN} selects neighbor nodes based on the inverse category frequency, and employs category-boosted negative sampling and adversarial learning in obtaining node embeddings. DDGraph~\cite{Ye2021DDGraph} utilizes dynamic user-item graph construction to select diverse items dynamically. A quantile progressive candidate selection scheme is used. DGRec~\cite{Yang2023DGRec} applies a submodular neighbor selection for diverse item acquisition, and a loss reweighting method to focus on long-tail items. However, these solutions have yet to consider user-item non-interaction diversity explicitly, which is essential for alleviating the echo chamber/filter bubble effects.

\begin{figure}[tb]
	\centering
	\includegraphics[width=1.0\linewidth]{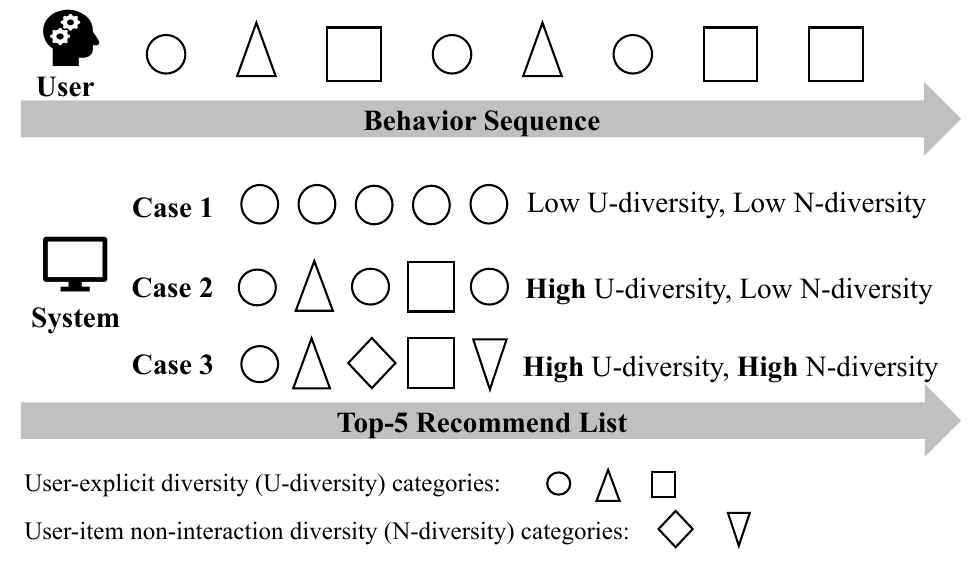}
	\caption{Illustration of U-diversity and N-diversity. Different shapes refer to different categories. Top: user historical behavior. Middle: three top-5 recommendation cases. Bottom: U-diversity categories and N-diversity categories.}
	\label{fig:undiversity}
\end{figure}

Category is used for grouping items. In e-commerce industry, the category groups products by considering their shared characteristics, e.g., apparel, food, home appliances, etc. In movie industry, category could be genre such as action, drama, and musical. There are a few of research on improving diversified recommendations by employing category information. Choi and Han~\cite{choi2010content} use the category correlation to improve diversity in collaborative filtering. DGCN~\cite{Zheng2021DGCN} applies category information in the category-boosted negative sampling scheme to obtain similar but negative items. It will help the model better to distinguish user's preferences within a category. Matroid Constraints ~\cite{abbassi2013diversity} relies on category similarity to measure the resemblance between candidate items and existing ones.
Besides, Hu and Pu~\cite{hu2011helping} propose an organization-based interface by grouping recommendations with categories. The results show that the organization interface effectively increases user’s perceived diversity of recommendations and further enhances their satisfaction.
Nevertheless, these works only treat categories as auxiliary information. Instead, we will use category to redefine diversity into U-diversity and N-diversity. Our proposed method CCDF directly models user's preferences in categories to obtain controllable diversity in order to achieve both high U-diversity and N-diversity.

\section{Method}
\subsection{Preliminaries}
\begin{table}[t]
	\centering
	\caption{The overall notations of our work.}
	\begin{tabular}{c|l}
		\toprule 
		Notation & Description \\
		\midrule
		$u$ & a user \\
		$i$ & an item \\
		$c$ & a category \\
		$\mathcal{U}$ & the set of users \\
		$\mathcal{I}$ & the set of items \\
            $\mathcal{I}_c$ & the set of items belongs to category c \\
            $\mathcal{C}$ & the set of categories \\
		$\mathcal{I}_u$ & the list of items interacted by user $u$ \\
		$\mathcal{P}_u$ & the basic profiles of user $u$ \\
		$\mathcal{F}_{uc}$ & the crossing features between $u$ and $c$\\
            $y$ & label, either positive or negative \\
		$s_{uc}$ & the number of interactions between $u$ and $c$ \\
		$\mathrm{C}(\cdot)$ & a mapping function for item to its category \\
		$c_{\text{neg}}$ & negative samples from user non-interacted category set\\
		$c_{\text{nei}}$ & neighbor samples from user interacted category set\\
            $n_{\text{neg}}$ & number of negative samples \\
            $n_{\text{nei}}$ & number of neighbor samples \\
            $\mathcal{G}$ & a heterogeneous graph of user-category interactions \\
            $\mathcal{V}$, $\mathcal{E}$ & the set of vertexes and edges in $\mathcal{G}$ \\
		\bottomrule
	\end{tabular}
	\label{table:notation}
\end{table}

\label{sec:diversity}
\subsubsection{Diversity}
Motivated by~\cite{Ye2021DDGraph}, we innovatively define diversity as two distinct definitions, user-explicit diversity (U-diversity) and user-item non-interaction diversity (N-diversity).
Specifically, U-diversity measures the extent to which the categories in recommended items overlap with the user's historical categories.
It represents the diversity in the set of recommended items that align with the user's previous interactions or expressed categories. N-diversity, also known as novelty, 
captures the degree to which the categories in recommended items differ from user historical behaviors or interests.
It focuses on introducing users to new and unfamiliar yet relevant categories, expanding their horizons, and encouraging exploration beyond user's existing preferences. 
Figure~\ref{fig:undiversity} shows three cases to illustrate the concepts of U-diversity and N-diversity in the context of e-commerce. Case 1 recommendation list only contains one category products which results in both low U-diversity and N-diversity. While case 2 recommendation list contains multiple categories, it fails to extend user's existing preferences. Thus it has high U-diversity but low N-diversity. Case 3 recommendation list not only has a good category converge on user's existing preferences, but also introduces novel items beyond user's known category preferences. As a result, it has both high U-diversity and N-diversity.
N-diversity is a key factor in alleviating echo chamber/filter bubble effects.
Achieving both high U-diversity and N-diversity will make real-world e-commerce recommender systems more effective and healthy.
Thus, we will evaluate the performance of our framework and state-of-the-art methods on both U-diversity and N-diversity tasks.

\subsubsection{Definitions}
Let $u\in \mathcal{U}$ denote the user and $i \in \mathcal{I}$ denote the item. $\mathrm{C}(\cdot)$ represents a mapping function that maps each item to its category,  and $c \in \mathcal{C}$ denotes the category.
An instance is represented by a tuple <$\mathcal{I}_u, \mathcal{P}_u, \mathcal{F}_{uc_t}, c_t,  s_{uc_t}, y$>,
where $\mathcal{I}_u$ is the user historical behaviors which are the items interacted by user $u$. $\mathcal{P}_u$ represents the basic profiles of user $u$, e.g., gender, age and so on. $\mathcal{F}_{uc_t}$ represents the crossing features between $u$ and $c_t$. $c_t$ is the target category, and $s_{uc_t}$ is the number of interactions between user $u$ and category $c_t$. $y$ represents the label, either positive or negative. Usually, if the user has interacted items with $c_t$, $y$ is positive. Negative examples will be sampled from user non-interacted category set, which is denoted as $c_{\text{neg}}$. The overall notations are summarized in Table \ref{table:notation}.


\subsection{The Proposed Framework}
As shown in Figure \ref{fig:motivation}, our proposed Controllable Category Diversity Framework (CCDF) is structured into two stages.
The first stage, User-Category Matching, is responsible for identifying a curated set of top-$K$ categories that align with the user's potential interests. The degree of control over category diversity is determined by the value of $K$. 
The second stage, Constrained Item Matching, focuses on retrieval of items within the top-$K$ categories chosen in the first stage. 

The advantage of our CCDF is that, it is able to explicitly control the diversity level of the recommended list by controlling parameter $K$. With a larger $K$, the retrieved items will be more N-diversified, which helps users discover new items, thereby alleviating echo chamber/filter bubble effects.

\begin{figure}[tb]
	\centering
	\includegraphics[width=1.0\linewidth]{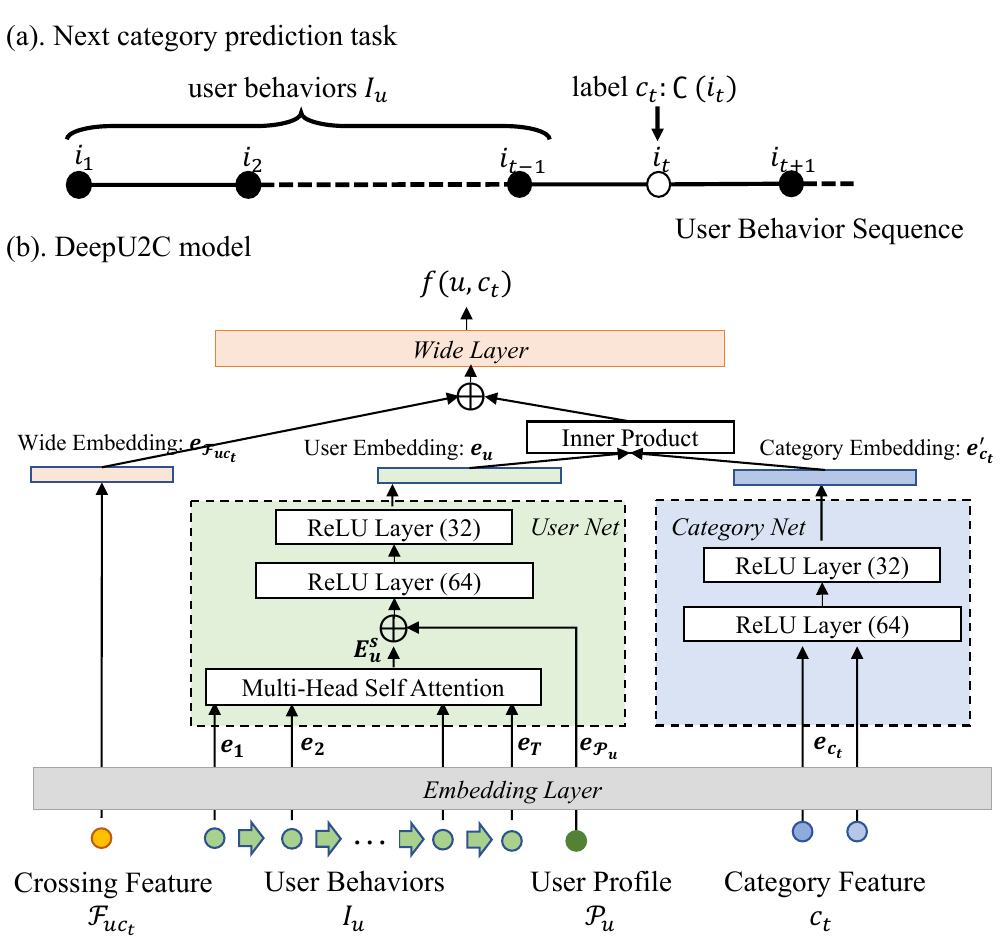}
	\caption{The User-Category Matching. (a): definition of next category prediction task. (b): DeepU2C model architecture.}
	\label{fig:architecture}
\end{figure}

\subsection{Stage I: User-Category Matching}
The goal of User-Category Matching in CCDF is to capture user's preferences in categories, and subsequently select the top-$K$ categories from the category candidate set that align with their interests to customize 
and control the category diversity level in recommendation list.
We model this stage as a next category prediction task, which aims to predict what the next category the user is most likely to click. When the target category $c_t$ exists in the user historical behaviors, i.e., $c_t \in \mathrm{C}(\mathcal{I}_u)$, it will be U-diversity next category prediction task, otherwise it will be N-diversity next category prediction task. We will evaluate our method and state-of-the-art baselines on these tasks.


\subsubsection{Next Category Prediction Task}
As shown in Figure~\ref{fig:architecture}~(a), we first obtain user historical behavior items in the form of a sequence in chronological order. Then we randomly choose an item from this sequence, whose category $c_t$  will be selected as the label. The user historical behaviors before this label will be selected as $\mathcal{I}_u$. The main objective of next category prediction task is to maximize the probability of user $u$ will click the item with category $c_t$.

\subsubsection{DeepU2C Model}
The basic idea of the model architecture is borrowed from Wide\&Deep network~\cite{Cheng2016WDL}. The motivation originates from its superior performance in memorization and generalization, which is quite suitable for improving U-diversity and N-diversity next category prediction tasks simultaneously. As shown in Figure~\ref{fig:architecture}~(b), the model mainly consists of \textit{Embedding Layer}, \textit{User Net}, \textit{Category Net}, and \textit{Wide Layer}.

There are four groups of input features, \textit{crossing features} $\mathcal{F}_{uc_t}$, \textit{user historical behaviors} $\mathcal{I}_u$, \textit{user profile} $\mathcal{P}_u$, and \textit{target category} $c_t$. 
Each feature is normally encoded into a high-dimensional one-hot vector and is further transformed into low-dimensional dense features by the \textit{Embedding Layer}. 
Thus \textit{crossing features}, \textit{user historical behaviors}, \textit{user profile} and \textit{target category} are transformed as $\mathbf{e}_{\mathcal{F}_{uc_t}}$, $\mathbf{E}_{\mathcal{I}_u}$, $\mathbf{e}_{\mathcal{P}_u}$ and $\mathbf{e}_{c_t}$ respectively. Note that, $\mathbf{E}_{\mathcal{I}_u}=\{\mathbf{e}_{1}, \mathbf{e}_{2}, ..., \mathbf{e}_{T}\} \in \mathbb{R}^{T \times {d}_{model} }$, where $T$ represents the length of user historical behaviors and ${d}_{model}$ is the dimension of item embedding $\mathbf{e}_{i}$.

The \textit{User Net} extracts user interests $\mathbf{e}_u \in \mathbb{R}^d$ from user historical behaviors $\mathbf{E}_{\mathcal{I}_u}$ and user profile features $\mathbf{e}_{\mathcal{P}_u}$. Multi-Head Self-Attention(MHSA)~\cite{Vaswani2017Attention} is used here to refine the behavior's item representation. Given the user historical behaviors $\mathbf{E}_{\mathcal{I}_u}$ as input, the MHSA outputs $\mathbf{E}_u^s \in \mathbb{R} ^{T \times d}$, which is formulated as:
\begin{equation}\label{eqn:mhsa}
\begin{aligned}
\mathbf{E}_u^s = \text{MHSA}(\mathbf{E}_{\mathcal{I}_u}) &= \text{Concat}(\textbf{head}_1, \textbf{head}_2, \cdots, \textbf{head}_H) \mathbf{W}^O, \\
\textbf{head}_h &= \text{Attention}(\mathbf{E}_{\mathcal{I}_u}\mathbf{W}_h^Q, \mathbf{E}_{\mathcal{I}_u}\mathbf{W}_h^K,
\mathbf{E}_{\mathcal{I}_u}\mathbf{W}_h^V) \\
 &= \text{Softmax} \Bigl(\frac{\mathbf{E}_{\mathcal{I}_u}\mathbf{W}_h^Q \cdot (\mathbf{E}_{\mathcal{I}_u}\mathbf{W}_h^K)^T}{\sqrt{d_h}} \Bigr) \cdot \mathbf{E}_{\mathcal{I}_u}\mathbf{W}_h^V,
\end{aligned}
\end{equation}
where $\mathbf{W}_h^Q ,\mathbf{W}_h^K, \mathbf{W}_h^V \in \mathbb{R}^{d_{model} \times d_h}$ are projection matrices of the $h$-th head for query, key and value respectively, The $H$ is the number of heads and $\mathbf{W}^O \in \mathbb{R}^{Hd_h \times d}$ is a linear matrix. Finally, a feed-forward network, consisting two fully-connected layers and ReLU activation~\cite{agarap2018deep}, is adopted to fusing $\mathbf{E}_u^s$ and user profile features $\mathbf{e}_{\mathcal{P}_u}$, 
\begin{equation}\label{eqn:usernet}
\begin{aligned}
\mathbf{e}_u = \text{ReLU}\Bigr(\text{ReLU}\bigr((\mathbf{e}_{\mathcal{P}_u} \oplus \text{sum}(\mathbf{E}_u^s))\mathbf{W}_1 + \mathbf{b}_1 \bigl)\mathbf{W}_2 + \mathbf{b}_2 \Bigl),
\end{aligned}
\end{equation}
where $\text{sum}(\cdot)$ is a sum-pooling operator and $\oplus$ means concatenation, $\mathbf{W}_1, \mathbf{W}_2$ and $\mathbf{b}_1, \mathbf{b}_2$ are learnable parameters of the feed-forward network.

The \textit{Category Net} is responsible for learning the latent embedding $\mathbf{e}_{c_t}^\prime \in \mathbb{R}^d$ of target category $c_t$. Here, we adopt two fully-connected layers with ReLU activation~\cite{agarap2018deep} to obtain a more expressive embedding for category, 
\begin{equation}\label{eqn:catenet}
\begin{aligned}
\mathbf{e}_{c_t}^\prime = \text{ReLU}\Bigr (\text{ReLU}(\mathbf{e}_{c_t} \mathbf{W}_3 + \mathbf{b}_3) \mathbf{W}_4 + \mathbf{b}_4\Bigl ),
\end{aligned}
\end{equation}
where $\mathbf{W}_3, \mathbf{W}_4$ and $\mathbf{b}_3, \mathbf{b}_4$ are learnable parameters.
It should be noted that these two deep networks, \textit{User Net} and \textit{Category Net}, are beneficial to the N-diversity next category prediction task due to their superior generalization performance.


The \textit{Wide Layer} is a generalized linear layer with the form $Wide(\mathcal{F}) = \mathbf{w}^T \mathcal{F}  + b$ where $\mathcal{F}$ consists of the crossing feature's embedding $\mathbf{e}_{\mathcal{F}_{uc_t}}$ and the dot product of user embedding and category embedding which are outputs from \textit{User Net} and \textit{Category Net} respectively. The \textit{Wide Layer} has a better memorization capability which is suitable to improving the performance in U-diversity next category prediction task.
In summary, after input <$\mathcal{F}_{uc_t}, \mathcal{I}_u, \mathcal{P}_u, c_t$>, we get the final user-category score:
\begin{equation}\label{eqn:nnetwork}
\begin{aligned}
f(u,c_t) &= \mathbf{w}^T \cdot \Bigl(\mathbf{e}_{\mathcal{F}_{uc_t}} \oplus Deep(\mathcal{I}_u, \mathcal{P}_u, c_t) \Bigr ) + b  \\
&= \mathbf{w}^T \cdot \Bigl(\mathbf{e}_{\mathcal{F}_{uc_t}} \oplus \langle \mathbf{e}_u, \mathbf{e}_{c_t}^\prime \rangle \Bigr ) + b,
\end{aligned}
\end{equation}
where \textit{Deep} contains \textit{Embedding Layer}, \textit{User Net} and \textit{Category Net}, $\mathbf{w}$ and $b$ are learnable parameters of \textit{Wide Layer}, and its output is a scalar.

\subsubsection{Combined Loss Function}
To improve the model performance, we employ a combined loss function consisting of cross entropy loss and triplet loss functions. The probability that a user $u$ will interact with an item with category $c_t$ is calculated as:
\begin{equation}\label{eqn:prob}
\begin{aligned}
\textbf{Pr}(c_t|\mathcal{I}_u, \mathcal{P}_u, \mathcal{F}_{uc_t}) = \frac{\exp(f(u,c_t))}{\sum_{c_j\in \mathcal{C}} \exp(f(u,c_j))},
\end{aligned}
\end{equation}
then we define the cross entropy loss $L_c$ as follows:
\begin{equation}\label{eqn:ce_loss}
\begin{aligned}
L_{c}(u,c_t) = - \log(\textbf{Pr}(c_t|\mathcal{I}_u, \mathcal{P}_u, \mathcal{F}_{uc_t})).
\end{aligned}
\end{equation}
The sum operator of Equation~\ref{eqn:prob} is computationally expensive. Thus, the sampled softmax technique~\cite{covington2016deep} with $n_{\text{neg}}$ negative categories generated by random sampling from user non-interacted category set  is adopted to train our model more time-efficiently.

In addition, we introduce the triplet loss~\cite{10.1007/978-3-319-24261-3_7} to enhance the ranking capability of our model. 
A heterogeneous graph $\mathcal{G}=(\mathcal{V}, \mathcal{E})$ is created based on user-category interactions,
where $\mathcal{V} = \mathcal{U} \cup \mathcal{C}$. There is an edge $e_{u,c}\in \mathcal{E}$ if $u$ has interacted with an item with category $c$. Then, for each positive sample data <$u, c_t$>, we sample $n_{\text{nei}}$ neighbor categories from $u$ as $c_{\text{nei}}$. Ultimately, we calculate the triplet loss $L_t$ as follows:
\begin{equation}\label{eqn:triplet_loss}
\begin{aligned}
L_t(u,c_t,c_{\text{nei}}) &= \frac{1}{n_{\text{nei}}}\sum_{c_j \in c_{\text{nei}}} \max \Bigl(\textbf{sign}(s_{uc_t} - s_{uc_j}) \\
&\cdot \bigl(f(u, c_j) - f(u, c_t) \bigr) + m, 0 \Bigr),
\end{aligned}
\end{equation}
where the margin $m$ is a hyper-parameter in triplet loss.
$\textbf{sign}(s) = 1$ when $s>0$, otherwise $\textbf{sign}(s) = -1$. Suppose we have <$u, c_j$>  where $s_{uc_t} > s_{uc_j}$, this loss function will steer our model's output to ensure that $f(u, c_t)$ is greater than $ f(u, c_j)$ by a margin $m$, thereby improving the ranking ability of our model. Note that $s_{uc_t}$ is the number of interactions between user $u$ and category $c_t$.

The overall loss function is a weighted sum of cross entropy and triplet loss functions,
\begin{equation}\label{eqn:overall_loss}
\begin{aligned}
L = \sum_{(u,c_t) \in \mathcal{D}}L_{c}(u, c_t) + \lambda \cdot L_t(u, c_t, c_\text{nei}),
\end{aligned}
\end{equation}
$\mathcal{D}$ is the collection of training samples, and $\lambda$ is a hyper-parameter.


\subsection{Stage II: Constrained Item Matching}
The objective of Constrained Item Matching in CCDF is to retrieve items accurately within the constraints of the top-$K$ categories which are output from previous stage. A popular mechanism in industry, posterior weighted score, is used here.
We calculate the real posterior click-through rate (CTR), conversion rate (CVR), and conversion purchase rate (CPR) for each item. The final score of each item is obtained by a weighted sum of CTR, CVR, and CPR, with weights 1, 10, and 100 respectively. These items are grouped by categories and the top-$N$ items of each category are retained. The above process is formulated as
\begin{equation}\label{eqn:c2i}
\begin{aligned}
\text{score}_i &= 1 \times \text{CTR}_i + 10 \times \text{CVR}_i + 100 \times \text{CPR}_i, \\
\mathcal{I}^{N}_c &= \underset{i \in \mathcal{I}_c}{\mathrm{argmax}} (\text{score}_i, N),
\end{aligned}
\end{equation}
where $\mathrm{argmax}(\text{score}_i, N)$ returns the top-$N$ items according to above calculated score. The top-$K$ categories in the first stage serve as triggers to retrieve items.

\section{Experiments}
In this section, we perform offline experiments and online A/B testing to validate the effectiveness of  CCDF. We evaluate the category diversity performance of our DeepU2C model by comparing it with these existing state-of-the-art diversity-oriented methods on offline real-world datasets. We also show the superiority of the whole CCDF framework by conducting real-world online A/B testing.

\subsection{Offline Experimental Setup}

\begin{table}[t]
	\centering
	\caption{Statistics of the offline datasets.}
	\begin{tabular}{llrrr}
		\toprule 
		&\small{Dataset} &\small{Users}  & \small{Categories}& \small{Interactions} \\
		\midrule 
		\multirow{2}{*}{\small{offline}} &
		\small{Taobao} & \small{50,000}  & \small{3,266} &\small{2,628,956} \\
		&\small{Alibaba.com} & \small{50,000}  & \small{4,360} & \small{3,419,462} \\
		\bottomrule 
	\end{tabular}
	\label{table:dataset}
\end{table}
\subsubsection{Datasets}
We evaluate the effectiveness of our method on category diversity performance with two real-world datasets. One of them is from an existing public e-commerce dataset, which is named as Taobao~\footnote{https://tianchi.aliyun.com/dataset/649}. The other is collected from our real-world industrial e-commerce website, which is referred as Alibaba.com~\footnote{https://www.alibaba.com/}.
Considering the evaluation of original large datasets is very time-consuming, we sample the top 50,000 active users according to interactions for both datasets.
We designate logs from the penultimate day as the validation set,  logs from the last day as the test set and the rest logs as the training set. 
The statistics of these two datasets are shown in Table~\ref{table:dataset}. 
The user historical behaviors are truncated with the length of 20 (i.e., $|\mathcal{I}_u|= 20$) for offline experiments and 30 for online A/B testing respectively. This truncation scheme is a common and reasonable practice for balancing complexity and accuracy.


\subsubsection{Baselines}
We compare our DeepU2C model with the following state-of-the-art methods.

\begin{itemize}
	\item \textbf{Statistics}: A naive statistical method obtaining the preferred categories by considering the user historical behavior timestamp and number of interactions with categories.
	\item \textbf{LightGCN}~\cite{DBLP:conf/sigir/0001DWLZ020}: A GCN-based model but removes the transformation matrix, non-linear activation, and self-loop.
	\item \textbf{DGCN}~\cite{Zheng2021DGCN}: A GNN-based diversified recommendation method with rebalanced neighbor discovering and category-boosted negative sampling.
	\item \textbf{DGRec}~\cite{Yang2023DGRec}: A variation of GNN-based diversified recommendation method.
\end{itemize}

\begin{table}[t]
	\setlength\tabcolsep{3.8pt} 
	\centering
	\caption{Offline experimental results on real-world datasets. The bold number is the best result, while the underlined number is the second best result.}
	\begin{threeparttable}
		\begin{tabular}{cccccccc}
			\toprule 
			&&\multicolumn{3}{c}{\small{Alibaba.com}} & \multicolumn{3}{c}{\small{Taobao}} \\
			\midrule 
			\multicolumn{2}{c}{\small{Method}} & \small{HR@5} & \small{HR@15} & \small{HR@30}& \small{HR@5} & \small{HR@15} & \small{HR@30}\\
			\midrule
			\multirow{5}{*}{\textit{\small{U}}\tnote{*}} &\small{Statistics} &\underline{\small{0.601}} &\small{0.718} & \small{$\boldsymbol{1.000}$} & \small{0.500}  &  \underline{\small{0.755}} & \small{$\boldsymbol{1.000}$} \\
			&\small{LightGCN} & \small{0.523} &\underline{\small{0.733}}&\small{0.833} &\underline{\small{0.525}}& \small{0.731}&	\small{0.829} \\
			&\small{DGCN}  &\small{0.449}&\small{0.674}&\small{0.790}&\small{0.448}&	\small{0.669}&	\small{0.783}\\
			&\small{DGRec} &\small{0.375}&\small{0.577}&\small{0.720}&\small{0.390}&	\small{0.587}&	\small{0.701}\\
			&\textbf{\small{DeepU2C}} &\small{$\boldsymbol{0.688}$}&\small{$\boldsymbol{0.810}$}&\small{$\underline{\small{0.843}}$}&\small{$\boldsymbol{0.781}$}&\small{$\boldsymbol{0.843}$}&\underline{\small{0.875}} \\
			\midrule
			\multirow{5}{*}{\textit{\small{N}}\tnote{*}} &\small{Statistics} & \small{0.000} & \small{0.000} & \small{0.000} & \small{0.000} & \small{0.000} & \small{0.000} \\
			&\small{LightGCN} &\small{0.088}&\small{0.228}&\small{0.368}&\small{0.061}&\small{0.149}&\small{0.255} \\
			&\small{DGCN} &\underline{\small{0.107}}&\small{0.241}& \small{0.372}& \small{0.068}&\small{0.156}&	\small{0.264}\\
			&\small{DGRec} &\small{0.098}&\underline{\small{0.256}}&\underline{\small{0.380}}& \underline{\small{0.085}}&\underline{\small{0.168}}&\underline{\small{0.278}} \\
			&\textbf{\small{DeepU2C}} &\small{$\boldsymbol{0.245}$} & \small{$\boldsymbol{0.458}$} & \small{$\boldsymbol{0.610}$} &\small{$\boldsymbol{0.248}$}&\small{$\boldsymbol{0.429}$}&\small{$\boldsymbol{0.580}$}\\
			\bottomrule 
		\end{tabular}
		\begin{tablenotes}
			\item[*] \small{\textit{U} means U-diversity task, \textit{N} means N-diveristy task.}
		\end{tablenotes}
	\end{threeparttable}
	\label{table:offline}
 \vspace{-1.3em}
\end{table}
\subsubsection{Evaluation Metric}
\textit{Hit Ratio (HR)} is used to measure the category accuracy of next category prediction task, defined as:
\begin{equation}
\begin{aligned}
HR@K = \frac{\sum_{(u,c_t) \in \mathcal{D}_{test}} \sigma(\text{target category} \mkern6mu c_t \mkern6mu  \text{in top-}K)}{|\mathcal{D}_{test}|},
\end{aligned}
\end{equation}
where $\sigma(\cdot)$ denotes indicator function and $\mathcal{D}_{test}$ denotes the test set consisting of pairs of user and target category ($u$, $c_t$). To evaluate all methods comprehensively, 
we divide $\mathcal{D}_{test}$ into two parts for U- and N-diversity tasks based on whether $c_t \in \mathrm{C}(\mathcal{I}_u)$ or not.

\subsubsection{Parameter Setting}
For parameter setting, we tune all methods using the validation set. Adam~\cite{DBLP:conf/iclr/ReddiKK18} is used as optimizer with a mini-batch size of $100$. The learning rate is fixed as $0.001$. 
The embedding sizes of input features, $\mathbf{E}_{\mathcal{I}_u}$, 
$\mathbf{e}_{c_t}$,
$\mathbf{e}_{\mathcal{F}_{uc_t}}$, and $\mathbf{e}_{\mathcal{P}_u}$, are set to $64$, $32$, $8$, and $8$, respectively. The number of heads $H$ used in MHSA is set to 8, and the fully-connected layers in \textit{User Net} and \textit{Category Net} employ hidden layers of size $64$ and $32$.
We randomly sample $500$ (i.e., $n_{\text{neg}}=500$) negative categories and $5$ (i.e., $n_{\text{nei}}=5$) neighbor categories for each positive sample. 
Other hyper-parameters $m$ and $\lambda$ in the loss function are tuned by Grid Search with $[0.2,0.3,0.4,0.5]$ and $[0.1,0.5,1.0,1.5,2.0]$ respectively.
To achieve the best results, the hyper-parameter $m$ is set as $0.4$ and $\lambda$ is set as $1.0$. We stop training until convergence within 5 epochs. 
\begin{table}[tb]
	\centering
	\caption{Offline ablation experimental results on Alibaba.com.}
	\begin{tabular}{clccc}
		\toprule 
		\multicolumn{2}{c}{\small{Method}} & \small{HR@5} & \small{HR@15} & \small{HR@30}\\
		\midrule 
		\multirow{4}{*}{\textit{\small{U}}} &\textbf{\small{DeepU2C}}&\small{$\boldsymbol{{0.688}}$}&\small{$\boldsymbol{0.810}$}&\small{$\boldsymbol{0.843}$} \\
		&\small{\qquad w/o triplet loss} &\small{0.645}&\small{0.782}&\small{0.810} \\
		&\small{\qquad w/o wide} &\small{0.483}& \small{0.602}& \small{0.668} \\
		&\small{\qquad w/o deep} &\small{0.534} &\small{0.689} &\small{0.788} \\
		\midrule 
		\multirow{4}{*}{\textit{\small{N}}} &\textbf{\small{DeepU2C}} & \small{$\boldsymbol{0.245}$}& \small{$\boldsymbol{0.458}$}& \small{$\boldsymbol{0.610}$} \\
		&\small{\qquad w/o triplet loss} &\small{0.221} &\small{0.387} &\small{0.552}  \\
		&\small{\qquad w/o wide} &\small{0.233} &\small{0.435} &\small{0.578} \\
		&\small{\qquad w/o deep} &\small{0.000} &\small{0.000} &\small{0.000} \\
		\bottomrule 
	\end{tabular}
	\label{table:ablation2}
\end{table}

\subsection{Offline Experimental Results}
The offline experimental results are shown in Table~\ref{table:offline}. Regarding U-diversity task, where the next category has occurred in user historical behaviors, our proposed DeepU2C achieves the best values of HR@5 and HR@15 for both Alibaba.com and Taobao datasets. LightGCN and Statistics are strong competitors. 
For the N-diversity task, where the next category does not occur in user historical behaviors, 
DGCN and DGRec obtain relatively better performance compared to their performance in U-diversity task. The reason is that they try to aggregate diverse categories in the graph. Nevertheless, 
our proposed method achieves $0.245$, $0.458$, and $0.610$ for HR@5, HR@15, and HR@30 on Alibaba.com respectively, ranking first place among various competitors. Similar results are observed on Taobao dataset. These results show the better generalization performance of the DeepU2C model, especially for non-interacted categories. One of the reasons can be attributed to the effectiveness of deep networks including \textit{User Net} and \textit{Category Net}, and the elaborated combined loss function.


\begin{figure}[tb]
	\centering
	\includegraphics[width=1.0\linewidth]{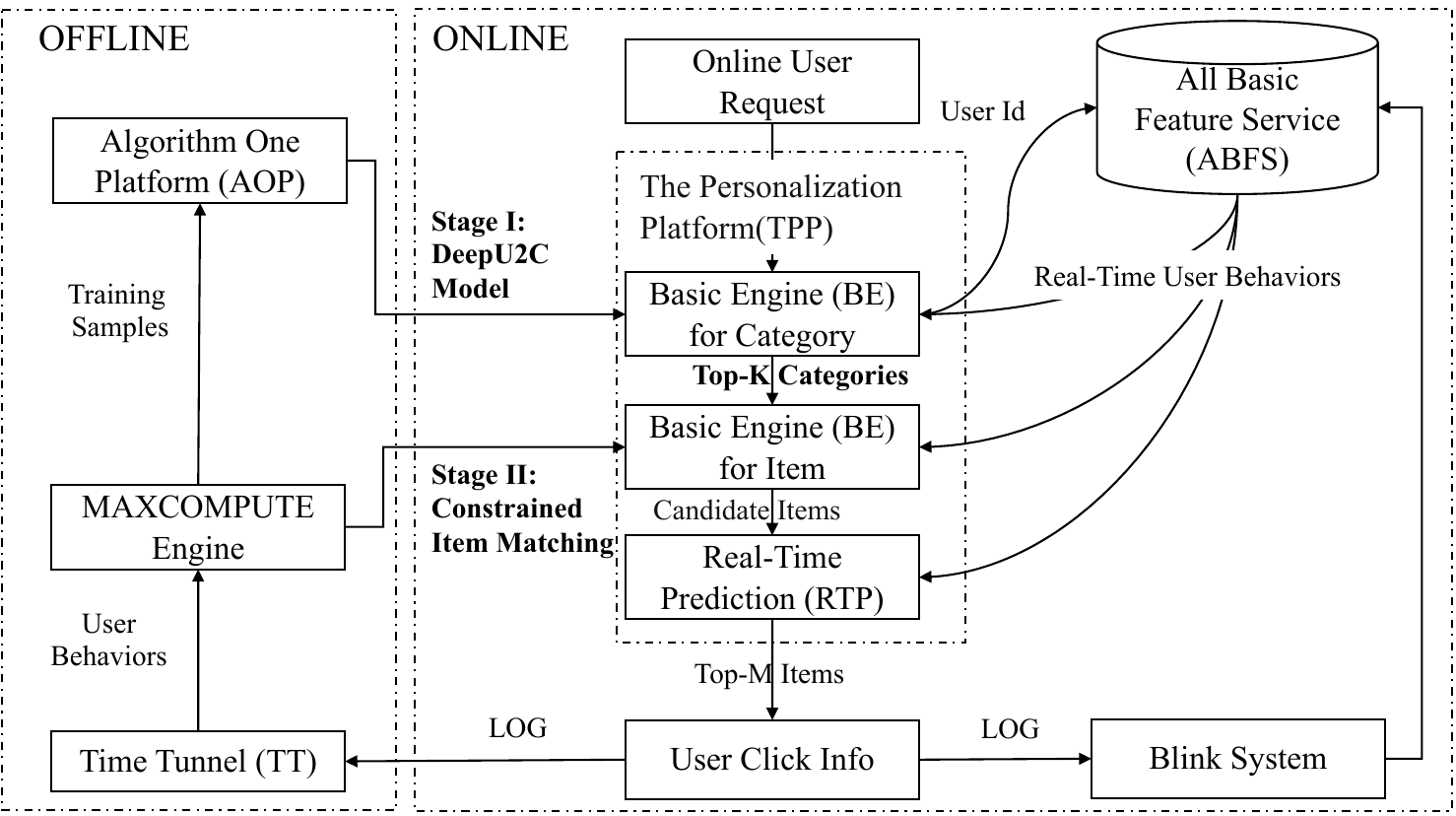}
	\caption{The online implementation architecture of our proposed CCDF.}
	\label{fig:serving}
\end{figure}
The ablation experimental results on Alibaba.com are shown in Table~\ref{table:ablation2}. We observe that \textit{Wide Layer} greatly improves the performance of DeepU2C on U-diversity task due to their excellent memorization capability. For example, after being equipped with the \textit{Wide Layer}, the value of HR@5 is increased from $0.483$ to $0.688$ on U-diversity task.
For N-diversity task, deep networks in \textit{User Net} and \textit{Category Net} which have a better generalization capability play a crucial role. Thus, the values of HR@5, HR@15 and HR@30 are dropped to zero after removing them. While the \textit{Wide Layer} has little effect on these metrics. Note that the triplet loss has positive effects on both two tasks.

\begin{figure*}[tb]
	\centering
	\includegraphics[trim = 0mm 0mm 0mm 0mm, clip, width=0.85\linewidth]{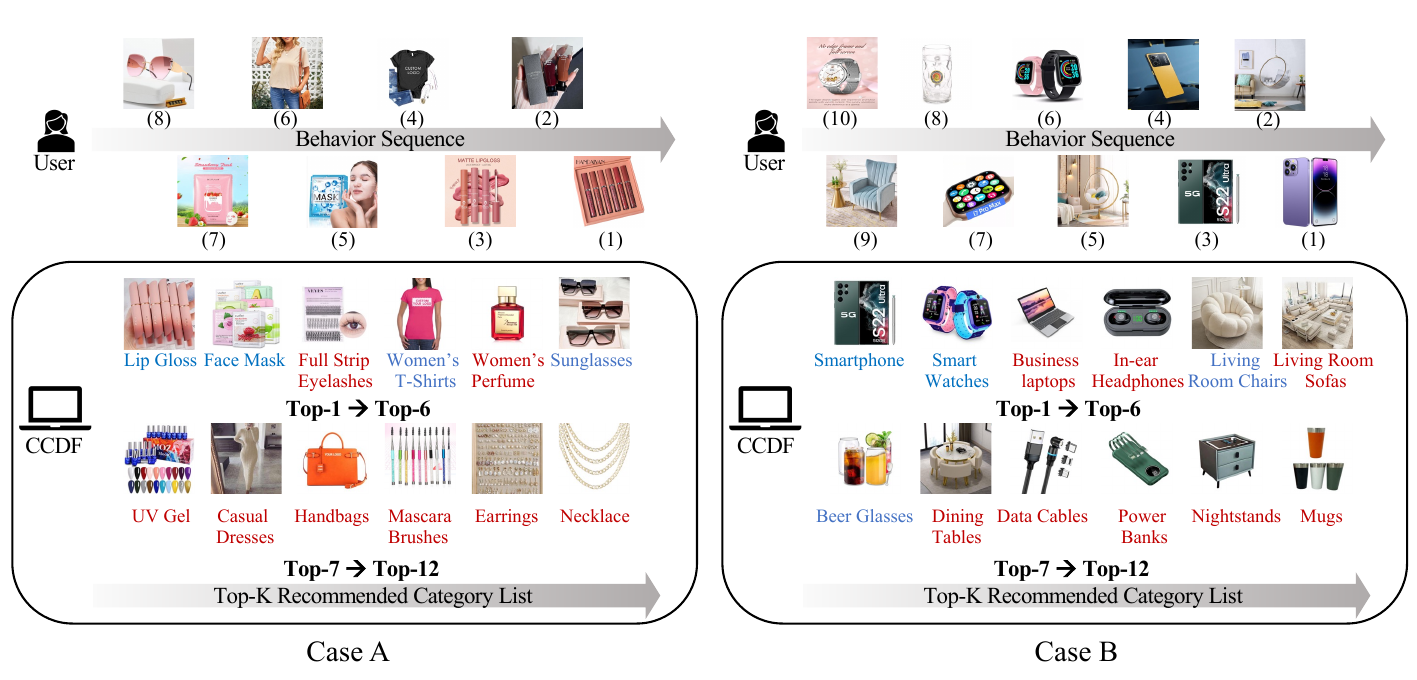}
	\caption{Two real-world cases from online traffic. The upper part is user historical behaviors in chronological order, while the lower part is the recommended categories from CCDF. U-diversity categories are in blue and N-diversity categories are in red. }
	\label{fig:case}
\end{figure*}
\subsection{Online Implementation Architecture}
We have deployed our proposed CCDF in a real-world production environment to do online A/B testing. Specifically, CCDF is deployed in Alibaba.com homepage recommendation scene, \textit{Just for You}, by leveraging several algorithm platforms in Alibaba Group. Figure \ref{fig:serving} illustrates the online architecture of our proposed CCDF. There are two main parts in the architecture, online and offline parts.

The online part is responsible for generating the final top-$M$ items that will be exposed to the end user. To be more specific, the online user request is served by The Personalization Platform (TPP) where the match and rank modules are inside. The DeepU2C model is deployed in Basic Engine (BE) for Category. To make the online serving more time-efficient, the category embeddings from \textit{Category Net} are calculated offline beforehand. We only calculate the remaining parts in real-time. In practice, the response time is usually less than 2ms for thousands of candidate categories. The Constrained Item Matching is deployed in BE for Item. It accepts the top-$K$ categories for item retrieval. Note that this match module has also deployed other candidate generation methods, such as item-based Collaborative Filtering method and deep candidate generation models. The next module is Real-Time Prediction (RTP), which is responsible for calculating each candidate item's score elaborately. Then the final top-$M$ items will be exposed to the end user. All Basic Feature Service (ABFS) is utilized here to return necessary user features, such as user profile features, real-time user historical behaviors, etc.

For the offline part, the user logs are recorded by Time Tunnel (TT). A big data platform, called MAXCOMPUTE, is used to process the logs in order to prepare training samples for training DeepU2C model. Algorithm One Platform (AOP) will consume these training samples to train the DeepU2C model. Once the training is finished, the model will be pushed to BE for Category for online serving. Note that, the posterior weighted score for each item in Constrained Item Matching is calculated in MAXCOMPUTE. Then $\mathcal{I}^{N}_c$ is implemented as an inverted index in BE for Item. $N$ is set as $300$ for online experiments.


\begin{figure}[tb]
	\centering
	\includegraphics[trim = 0mm 0mm 0mm 0mm, clip, width=1.0\linewidth]{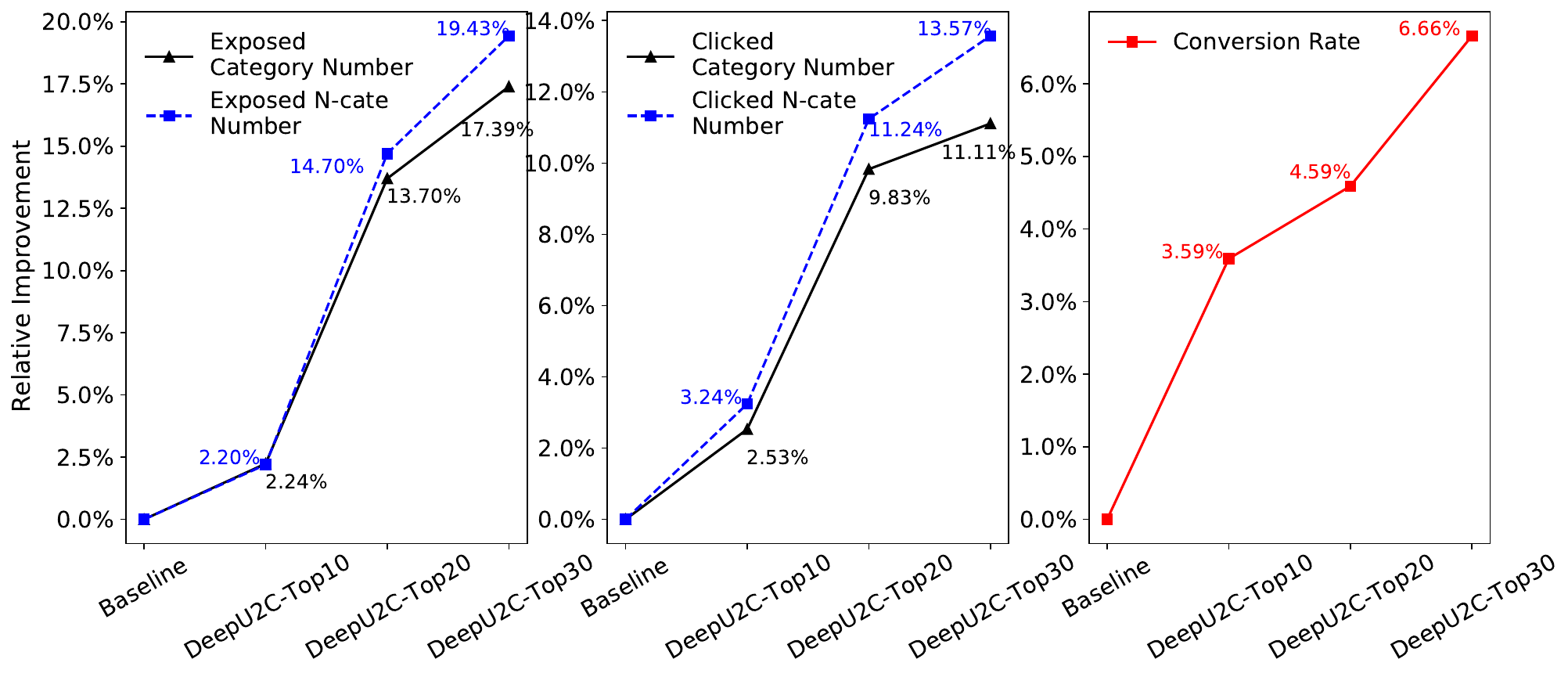}
	\caption{The improvements of CCDF with different top-$K$ parameters on online A/B testing compared with the baseline. $K$ is the controllable parameter for diversity which is set as 10, 20, and 30.}
	\label{fig:abtest}
 \vspace{-1.2em}
\end{figure}

\subsection{Online A/B Testing}
Usually, the efforts to deploy a new model in the production environment are not trivial, thus we compare our proposed DeepU2C, which has obtained the best performance in offline experiments, with existing online baseline Statistics method for two weeks. Both of them are employed with the same Constrained Item Matching, i.e., posterior weighted score. 
To make comparisons fairly, all of other parts are untouched.
For training  online version of DeepU2C, we collect the 90-day of real online user interactions from our platform, which has millions of users and thousands of categories.
\subsubsection{Numerical Results}
\label{sec:number_result}
The A/B testing results are presented in Figure~\ref{fig:abtest}. When $K=10$, the exposed category number and clicked category number are increased by $2.24\%$ and $2.53\%$ respectively. Furthermore, the corresponding conversion rate is increased by $3.59\%$, which proves the assumption that diversity is able to improve user's experiences and increase website revenues. When we tweak the $K$ to $20$ and further to $30$,  the exposed category number is increased by $13.70\%$ and $17.39\%$, the clicked category number is increased by $9.83\%$ and $11.11\%$. The conversion rate is increased by $4.59\%$ and $6.66\%$.
Meanwhile, it is important to report the exposed and clicked N-category number which are used to denote the N-diversity. These values indicate how many categories that have not occurred in user historical behaviors are exposed and clicked. Under $K$ in settings $10$, $20$, and $30$, the exposed N-category number and clicked N-category number are increased by $2.20\%$, $14.70\%$, $19.43\%$ and $3.24\%$, $11.24\%$, $13.57\%$ respectively.
All these improvements are statistically significant using an unpaired t-test.
These findings provide clear validation of the superiority of our proposed CCDF. 
With an increase in $K$, not only does the diversity level improve, but also there is a significant enhancement in conversion rate.
The continuous improvements of N-diversified metrics also indicate that our framework is able to help user to obtain new products and discover new needs effectively, and then alleviate the echo chamber/filtering bubble effects of our system.
Notably, the retrieval of  N-diversified items, which are neglected by other candidate generation methods in match module, is the key reason why our framework is able to achieve huge improvements on conversion rate online.

\subsubsection{Case Study}
We present two real-world cases from online traffic in Figure~ \ref{fig:case}. User historical behaviors are listed in the upper part, while the recommended categories from CCDF are shown in the lower part. For case A, the user has expressed strong interests in \textit{Lip Gloss} by clicking products (1), (2), and (3), and \textit{Face Mask} by clicking products (5) and (7). Meanwhile, this user has some sparse interests, \textit{Women's T-Shirts} and \textit{Sunglasses}. Our proposed CCDF successfully predicts the user's preferences for these four categories, which would result in a high U-diversity. Furthermore, when the $K$ is increased, we find that more N-diversity categories have been presented. These categories are relevant to the user's preferences to some extent, which broadens user's interests. For example, the new categories \textit{Full  Strip Eyelashes}, \textit{Women
s Perfume}, \textit{UV Gel} and \textit{Mascara Brushes} are extended from user original interests \textit{Lip Gloss} and \textit{Face Mask}. Besides categories in cosmetics, there is a high probability that a lady would like to view products in \textit{Casual Dresses}, \textit{Handbags}, \textit{Earrings}, and \textit{Necklace}. The same conclusions can also be observed in case B. \textit{Smartphone} is extended to \textit{Data Cables} and \textit{Business laptops}, \textit{Living Room Chairs} is extended to \textit{Living Room Sofas} and \textit{Dining Tables}, etc. Both cases have verified the effectiveness of CCDF, which is able to achieve high U-diversity and N-diversity by tweaking the value of $K$. 

\begin{figure}[tb]
	\centering
	\includegraphics[trim = 0mm 0mm 0mm 0mm, clip, width=1.0\linewidth]{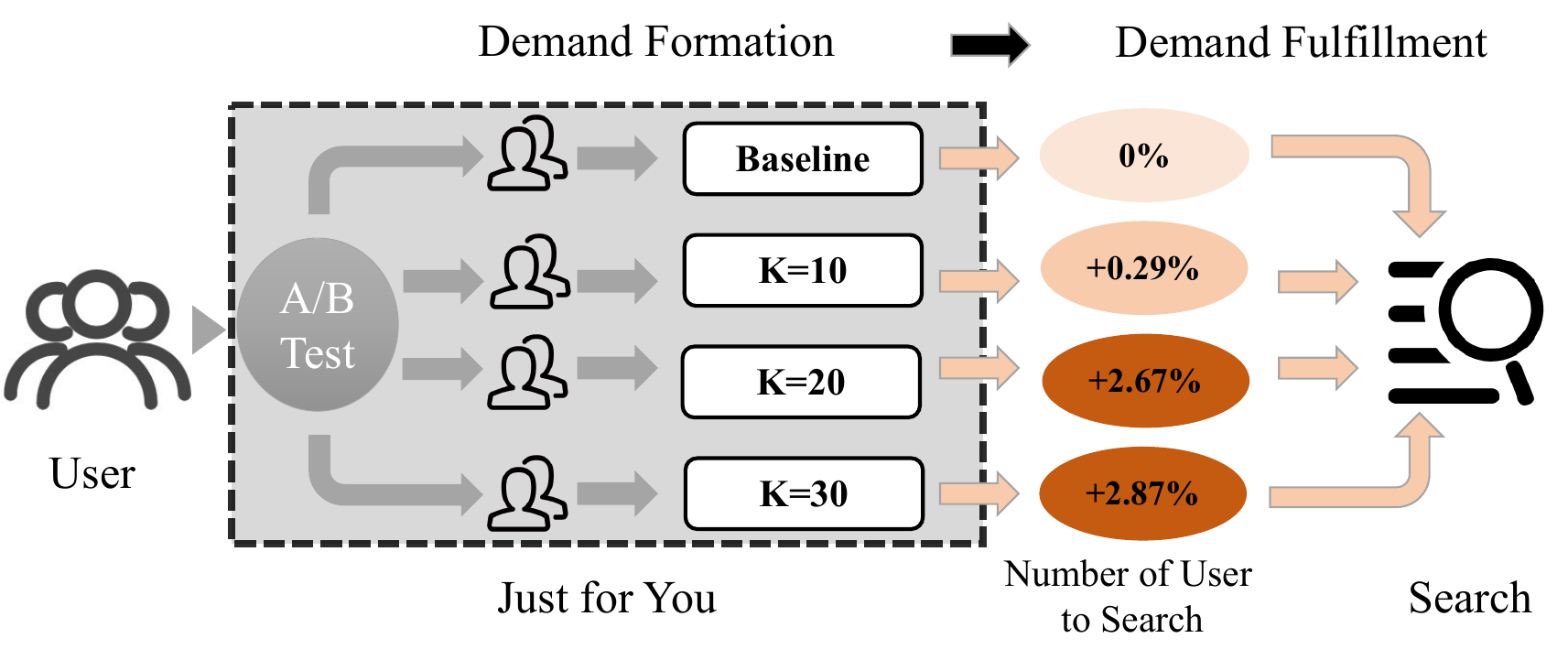}
	\caption{The demonstration of user flow from homepage recommendation scene, Just for You, to search.}
	\label{fig:search}
 \vspace{-1.2em}
\end{figure}

\subsubsection{The Impacts on User Search}
These online experiments on homepage recommendation scene,  \textit{Just for You}, also have positive impacts on user's search activities. 
Figure~\ref{fig:search} shows the user flow from \textit{Just for You} to search under different experimental settings. Table \ref{table:search} presents the experimental results of users from \textit{Just for You} to search.
We observe that when the value of $K$ is increased, the diversity-level of recommendation list is increased (as presented in Section~\ref{sec:number_result}), then more users in recommendation scenes are willing to go to search. The number of users from \textit{Just for You} to search is increased by $0.29\%$, $2.67\%$, and $2.87\%$ with $K$ setting as $10$, $20$, and $30$ respectively. One of the reasons is that the diversified recommendations help user to access more items, and in turn encourage user to go to search to refine their needs. The improved conversion category number and conversion rate in search further validate the promotional effect of diversified recommendations on user search. More specifically, in search, the conversion category number of these users are increased by $0.97\%$, $2.28\%$, and $4.01\%$ respectively, and the conversion rate of these users are improved by $2.28\%$, $3.57\%$, and $5.54\%$ respectively.
The above experimental results support the existence of \textit{complementary effects}~\cite{yuan2022recommendation} between recommendation and search.
Some customers may form new demands while browsing diverse results in recommendations (demand formation) and then they actively initiate a search to meet their needs more efficiently (demand fulfillment). 

\begin{table}[t]
	\centering
        \caption{Experimental results of users from \textit{Just for You} to search on online A/B testing.}
	\begin{tabular}{cccc}
		\toprule 
            \small{Method} & \small{\makecell{Number of User \\ to Search}} &\small{\makecell{Conversion Category \\ Number in Search}}  & \small{\makecell{Conversion Rate \\ in Search}} \\
            \midrule
		\small{Baseline} &\small{0\%} &\small{0\%}  & \small{0\%} \\
		\small{K=10} & \small{+0.29\%} & \small{+0.97\%}  & \small{+2.28\%}  \\
		\small{K=20} & \small{+2.67\%} & \small{+2.28\%}  & \small{+3.57\%}  \\
		\small{K=30} & \small{+2.87\%}& \small{+4.01\%}  & \small{+5.54\%} \\
		\bottomrule 
	\end{tabular}
	\label{table:search}
 \vspace{-1.3em}
\end{table}

\section{Conclusion}
In this paper, we innovatively define the diversity as two distinct definitions, user-explicit diversity (U-diversity) and user-item non-interaction diversity (N-diversity) based on user historical behaviors. Then, we propose a succinct and effective Controllable Category Diversity Framework (CCDF), which is able to explicitly control category diversity to achieve both high U-diversity and N-diversity. 
There are two stages in CCDF, User-Category Matching and Constrained Item Matching. 
We formalize User-Category Matching as a next category prediction task and construct a DeepU2C model which incorporates User Net, Category Net, Wide Layer, and a combined loss. 
This two-stage framework divides the responsibilities of category diversity and item accuracy into two different stages,  which equips CCDF with the controllability on category diversity by parameter $K$. The larger the parameter $K$, the more N-diversified items will be retrieved. As a result, it is beneficial to alleviate echo chamber/filter bubble effects. Offline experimental results on real-world datasets and online A/B testing results validate the superiority of our framework. The complementary effects between recommendation and search have been observed in our online A/B testing.


In the future, we will strive to improve the accuracy of DeepU2C model, especially for N-diversity task in stage I. Then we will try to model user's preferences for items under constraint of specific category through deep networks to improve the accuracy of stage II.
\bibliographystyle{ACM-Reference-Format}
\bibliography{u2c}

\end{document}